\documentclass[11pt]{article}
\input epsf

\title{\bf PT Symmetry and Hermithean Hamiltonian in the Local
Supercritical Pomeron Model}
\author{M.A.~Braun$^a$ and G.P.~Vacca$^b$\\
${}^a$ University of S.Petersburg,
198504 S.Petersburg, Russia\\
${}^b$INFN - Sez. di Bologna, Dip. di Fisica, Via Irnerio 46, 40126 Bologna, Italy }
\def\beq{\begin{equation}}
\def\eeq{\end{equation}}

\def\phid{\phi^{\dagger}}

\begin{document}
\maketitle
\medskip
\begin{center}
{\bf Abstract}
\end{center}
The local reggeon field theory is studied perturbatively
taking advantage of the PT symmetry in the Hamiltonian formulation. In the
lowest non trivial order we show that the pomeron interactions renormalize the slope.
In the same order we find a non local pair potential acting between pomerons,
which has a singular structure. However the analysis of the scattering
operator shows that at small coupling constant bound states do not appear so
that the two-particle spectrum is not changed.

\section{Introduction}
In  recent years the study of strong interactions in the so called Regge  kinematics has
experienced a rebirth of interest due to advances in
the Quantum Chromodynamics studies, originated from the seminal work about the
BFKL ~\cite{BFKL} pomeron. Subsequent investigations have unveiled the basic
effective interaction~\cite{BW,Mueller,BV} of  propagating BFKL pomerons. Equations to resum
tree diagrams  at large $N_c$ were derived for onium-nucleus scattering ~\cite{Bal,Kov}  (BK
equation) and nucleus-nucleus scattering ~\cite{B1}.An effective quantum field theory
describing the pomeron interaction at large $N_c$ was constructed in ~\cite{BV,B2}, which
in principle allows to find quantum corrections related to pomeron loop diagrams.
Similar  studies were conducted in the alternative  dipole and JIMWLK techniques,
 where also attempts to sum pomeron loops were made under certain drastic 
approximations ~\cite{CGS}.

Pomeron loops play a secondary role in the scattering with nuclei at not too high energies.
But for the proton scattering and at  asymptotic energies their contribution
cannot be neglected. Summing pomeron loops is a formidable task. So, as a first step,
it is worthwhile to study some features of a much
simpler quantum field theory of the local supercritical pomerons,
introduced by V.N.Gribov many years ago to sum reggeon diagrams which describe
interacting pomerons in the phenomenogical approach. This local reggeon field theory
(LRFT) possesses many features similar to  the QCD pomeron theory. In particular,
for the supercritical pomeron with the intercept $\alpha(0)>1$, it is
also non-perturbative at high energies, so that loop contributions have to be
summed by some technique, which may prove to be useful also for the QCD.
Some very beautiful results in this direction were obtained for the even simpler 
LRFT living in zero transverse dimension (``a toy model'), which in fact
reduces  to the quantum mechanics. In particular it was shown that loops indeed play 
a decisive role at high energies and essentially transform the initially supercritical
pomeron into a weakly subcritical. These results were in fact obtained long ago
~\cite{schwimmer,acj} and were recently re-analyzed in ~\cite{KL,Bond,BVtoy}.
Unfortunately it is not straightforward to extend these findings to the realistic
LRFT living in two transverse dimensions. Differentg approximationshave  lead
different authors to either predict a phase transition or claim  
complete inconsistency of the model.
More studies are needed to clarify the situation.

In this paper we draw attention to the fact that in the Hamiltonian approach
LRFT belongs to the class of models with a non-Hermitean Hamiltonian, which
possesses a certain symmetry equivalent to the $PT$ symmetry in the ordinary
quantum mechanics.
This was noted  in ~\cite{BVtoy} for the toy model in zero transverse
dimension but remains true also for the realistic case. The $PT$-symmetric 
non-Hermithean Hamiltonians have been extensively studied for some time
~\cite{bender,mostafazadeh} and
we are going to apply some of the techniques developed in this study for the
LRFT with a supercritical pomeron. In particular using the $PT$ invariance
we demonstrate that the non-Hermitean Hamiltonian of LRFT has only real eigenvalues 
and transform it to an Hermthean one by a suitable
similarity transformation. As for zero transverse dimensions in  ~\cite{BVtoy},
we are able to do this only by perturbation expansion in the coupling constant
of the pomeron interaction $\lambda$. This does not allow us to find the true asymptotic 
of the amplitudes in LRFT at high energies with all loops taken into account
but gives results valid up to rapidities $y\sim 1/\lambda^{2N}$ when terms up to
order $\lambda^{2N}$ are included. Thus, at small $\lambda$, 
higher orders in the perturbation expansion of the Hamiltonian allow to study 
higher and higher rapidities.

In the course of our study we have to perform renormalization of the loop correction to
the pomeron intercept, which is divergent in the lowest order, as  is well known.
As a result we find that the slope of the pomeron trajectory is decreased by interaction.
We find that this is the only change in the supercritical pomeron spectrum at order 
$\lambda^2$.  

The plan of the paper is as follows.
In the next section we briefly review some features of $PT$ symmetric quantum
systems relevant to our study. Then we introduce the LRFT model and its
Hamiltonian formulation and discuss The $PT$ and $CPT$ symmetries of the Hamiltonian.
Next two chapters are devoted to the calculation of the metric operator
and the Hermithean Hamiltonian in the lowest non-trivial order.
In the fifth section we study and renormalize the single particle term of the
Hamiltonian and derive the effective two particle non-local potential.
In the sixth section we study the scattering states associated with this
potential and show that no bound states are present at perturbative
level.
We then draw our conclusions.
Some more technical details have been added in a few appendices.

\section{PT symmetric Quantum Mechanics and QFT}
Since it was noted~\cite{bender} that there exist non hermithean Hamiltonians $H$
having a real spectrum bounded from below, provided boundary conditions for the wave
functions of the associated Sturm-Liuville differential problem are properly
defined, a lot of investigations have been done.

An important observation has been that a class of such Hamiltonians possesses an unbroken $PT$
symmetry, so that eigenstates of $H$ can be chosen to be eigenstates of
$PT$. After investigating the scalar product for the state space it has been found
that there is a natural choice $(f,g)=\int dx [PT f]^t(x) g(x)$, $t$ denoting the
transpose operation, which gives an associated norm conserved in time.
The problem of this choice is that it is not leading to an
Hilbert space with a positive norm. The states are splitted in two classes
with positive and negative $PT$-norms. The good news has been that it is generally
 possible to find a new
symmetry operator, denoted by $C$, whose eigenvalues are precisely the sign of the 
$PT$-norm.

Having $[C,PT]=0$ and $[C,H]=0$, one can define a new scalar product by
$\langle f|g \rangle=\int_\Gamma dx [CPT f]^t(x) g(x)$, which leads
to a positive norm conserved in time and
therefore to a physically acceptable probabilistic interpretation.
In such a case one can also define observables to be operators such that
$O^t=CPT O CPT$, relation which coincides with the
usual hermiticity condition in the conventional quantum mechanics where $C=P$.
We stress that the main conceptual point here has been that if one insists to define a
quantum theory with conserved probability using a $PT$-symmetric non-Hermithean Hamiltonian
one has to use the new mathematically well defined $CPT$-scalar product, which
depends on the Hamiltonian itself.

On the other hand in many physical situations the scalar product of the wave functions  
is well
defined from the start and cannot be changed. With this scalar product the probabilities
given by the norm of the wave functions will not be conserved in time. However this does not
present any difficulty, since either they are not conserved physically, as
for decaying channels, or evolution in fact goes not in time but in rapidity,
as in LRFT, when requirement that the norm be conserved is absent.
Thus the new norm introduced in earlier studies is not relevant for LRFT. However
other points have a direct application.

In the next section we shall find that the Hamiltonian of the LRFT indeed posseses the PT
symmetry and has real eigenvalues. 
Construction of the new  symmetry operator $C$ depends on the  Hamiltonian.
If one knows eigenstates of $H$ then one can use them to explicitly find a
representation of $C$.
A  more convenient way, which admits a perturbative approach~\cite{BenderQFT}, is the following.
Consider a system whose Hamiltonian has the form 
\beq 
H=H_0+\lambda H_I \,,
\eeq 
where the free part is given by a Hermithean $H_0$ and the interaction part by an anti-Hermithean $H_I$.
Define the parity operator $P$ with $P^2=1$ to transform $H$ into $H^\dagger$, which implies
\beq
[H_0,P]=0,\ \  \{H_I,P\}=0,\ \ P^2=1.
\eeq
Now one looks for the symmetry operator $C$
satifying
\beq
[C,H]=[C,PT]=0
\label{chpt}
\eeq
 in the form
\beq
 C=e^Q P
\eeq
where $Q$ is an Hermithean operator.
From (\ref{chpt}) we find a relation~\cite{BenderQFT} 
\beq
2\lambda \, e^Q H_I=[e^Q,H]\,,
\eeq 
which can be solved perturbatively, using the expansion $Q=\lambda
Q_1+\lambda^3 Q_3+...$, by requiring
\beq
 [H_0,Q_1]=-2H_I,\ \
[H_0,Q_3]=-\frac{1}{6}\Big[[H_I,Q_1]Q_1\Big] \label{condq}
\eeq
and so on. 

The operator $e^Q$ can be used to define a similarity transformation which
maps the PT-symmetric Hamiltonian $H$ onto an Hermithean Hamiltonian $h$ with the
same set of eigenvalues.
It is easy to show that also
\beq 
e^{-Q} H e^Q=H^\dagger.
\label{eqh} 
\eeq
Indeed we have
$ e^QPH-He^QP=0$
Multiplying by $P$
and using $PHP=H^\dagger$ we find (\ref{eqh}). 
As a result, we find
\beq
 h=e^{-Q/2} He^{Q/2}=e^{Q/2} H^\dagger e^{-Q/2}=h^\dagger,
\label{defh}
\eeq
so that $h$ is the equivalent Hermithean Hamiltonian.
Once $Q$ is known as a power series in $\lambda$ the Hamiltonian $h$
can also be found in the same form:
$h=h^{(0)}+\lambda^2h^{(2)}+\lambda^4h^{(4)}+...$ where
\[ h^{(0)}=H_0,\]
\[h^{(2)}=\frac{1}{4}[H_I,Q_1],\]
\beq
h^{(4)}=\frac{1}{4}[H_I,Q_3]+\frac{1}{32}\Big[[H_0,Q_3]Q_1\Big]
\label{hamilh}
\eeq
and so on.

In the next sections we show that it is possible to apply these general
results to the LRFT
and find the symmetry operator $Q$ and the Hermithean Hamiltonian $h$ 
in the first non-trivial order of the  perturbation theory.

\section{PT symmetry and operator $Q$ in the LRFT}
The LRFT can be defined as a theory of two fields $\phi(y,x)$ and $\phid(y,x)$
depending on rapidity $y$ and transverse coordinates $x$ with a 
Lagrangian density
\beq
{\cal L}=\phid(\partial_y-\mu-\alpha'\nabla_x^2)\phi+i\lambda
\phid(x)\Bigl[\phid(x)+\phi(x)\Bigr]\phi(x),
\eeq
where $\mu>0$ is the intercept minus unity and $\alpha'$ is the slope
of the pomeron trajectory. With $\mu>0$ the corresponding functional integral
is divergent and the only way to define the theory beyond the set of
perturbative Feynman diagram is the analytic continuation from $\mu<0$ when
the theory is well defined.
A constructive way to do this continuation is the Hamiltonian approach.
One sets up a quasi-Schroedinger equation for the wave function $\Psi$:
\beq
\frac{ d\Psi(y)}{dy}=-H\Psi(y),
\eeq
where the Hamiltonian has the form
\beq 
H=H_0+\lambda H_I
\eeq 
with the free part given by
\beq
 H_0=\int d^2x
(-\mu\phid(x)\phi(x)+\alpha'\nabla\phid(x)\nabla\phi(x)), 
\eeq 
the
interaction part 
\beq 
H_I=i\int
d^2x\, \phid(x)\Bigl[\phid(x)+\phi(x)\Bigr]\phi(x),
\eeq
and the standard commutation relations between $\phi$ and $\phid$:
\beq
[\phi(x),\phid(x')]=\delta^2(x-x').
\eeq
The scattering amplitude with the target ('initial') state $\Psi_i(y_1)$ at rapidity 
$y_1$
and the projectile ( 'final') state  $\Psi_f(y_2)$ at rapidity $y_2>y_1$ is defined as
\beq
iA_{fi}(y_2-y_1)=\langle \Psi_f(y_2)|e^{-H(y_2-y_1)}|\Psi_i(y_1)\rangle.
\label{ampl}
\eeq
One can demonstrate that the perturbation expansion in powers of $\lambda$ of this expression
reproduces the standard Reggeon diagrams of the LRFT and also that (\ref{ampl})
satisfies the requirement of symmetry between the target and projectile
(see ~\cite{BV})

The Hamiltonian is not
Hermithean and our first task is to demonstrate that its energy levels 
are all real. This is of course  trivially seen in the perturbation theory.
Since $H_0$ has its eigenstates with a fixed number $n$ of pomerons, the
energy change can only be accomplished by action
of an even number  of interactions $H_I$. However we can also prove it on more general grounds.
To this end we consider symmetry operations applied to $H$.
In correspondence with the definitions in the previous section we introduce parity $P$
as a transformation of the fields
\beq
\phi(y,x)\to -\phi(-x),\ \ \phid(y,x)\to -\phid(-x).
\eeq
It follows that indeed $PHP=H^{\dagger}$ and of course $P^2=1$.
Next we introduce the `time reflection' $T$ as taking the complex conjugate of 
all coefficient functions without changing the fields.

It is evident that $P$ and $T$ commute and that their product $PT$
will leave both parts of the Hamiltonian intact.
So we indeed find a $PT$ symmetry of the LRFT Hamiltonian 
\beq
[PT,H]=0.
\eeq
As a result, if an eigenfunction of $H$ is presented as the action of some operator
depending on $\phid$ on the vacuum $\Psi_0$, then it has to be of the form
\beq
F(i\phid)\Psi_0,\ \ {\rm with}\ \ F(z^*)=F^*(z),
\eeq
i.e. $F$ has to be a real function of its complex argument. The eigenvalue is then
\[
\langle\Psi|H|\Psi\rangle=\langle\Psi_0|F(-i\phi) H F(i \phid)|\psi_0\rangle
\]
and is obviously real, since the Hamiltonian can also be written as a real function
of $i\phid$ and $-i\phi$ and the total mumber of operators $\phi$ and $\phid$
 has obviously to be the same.

Following the technique discussed in the previous section our aim is to
find an appropriate $Q$-operator, which will allow us to pass to a Hermithean 
Hamiltonian $h$ by transformation (\ref{defh}).
Once we find the latter, the amplitude will be given by
\beq
iA_{fi}(y_2-y_1)=\langle e^{Q/2}\Psi_f(y_2)|e^{-h(y_2-y_1)}|e^{-Q/2}\Psi_i(y_1)\rangle.
\eeq
Thus evolution will be accomplished by the Hermithean $h$ and operators $e^{\pm Q/2}$
will transform (differently) the initial and final states. In particular
 the pomeron
Green function at rapidity $y$ and momentum $k$ will be given as as
\beq
\delta^2(k-k')G(y,k)=<0|\phi(k)e^{Q/2}e^{-yh}e^{-Q/2}\phid(k')|0>.
\eeq

To construct $Q$ we shall use perturbative equations (\ref{condq}) presented in the
previouss section.

In this paper we shall restrict ourselves to the first non-trivial order
in the triple pomeron coupling constant $\lambda$, that is constructing
$Q_1$ and $h^{(2)}$.
To accommodate to our notations in ~\cite{BV} we seek $Q_1$ in the form
\beq
Q_1=-2\frac{i}{\mu}\int d^2x_1d^2x_2d^2x_3
\Big(f_1(x_1,x_2,x_3)\phid_1\phi_2\phi_3
+f_2(x_1,x_2,x_3)\phid_1\phid_2\phi_3\Big)\,,
\eeq
where we denote $\phi_1\equiv\phi(x_1)$ etc.
To calculate the part of the commutator $[Q_1,H_0]$ which contains 
function $f_1$ we need to know 
\[
[\phid_1\phi_2\phi_3,\phid(x)\phi(x)]=
\delta^2(x_2-x)\phid_1\phi_3\phi(x)+\delta^2(x_3-x)\phid_1\phi_2\phi(x)
-\delta^2(x_1-x)\phid(x)\phi_2\phi_3\,
\]
and
\[
[\phid_1\phi_2\phi_3,\nabla\phid(x)\nabla\phi(x)]
\]\[=
\nabla\delta^2(x_2-x)\phid_1\phi_3\nabla\phi(x)+
\nabla\delta^2(x_3-x)\phid_1\phi_2\nabla\phi(x)
-\nabla\delta^2(x_1-x)\nabla\phid(x)\phi_2\phi_3\,,
\]
where $\nabla$ refers to differentiation in $x$.
So this  part of the commutator $[Q_1,H_0]$ takes the form
\beq
[Q_1^{(1)},H_0]=-2\frac{i}{\mu}
\int\prod_{i=1}^3d^2x_i\phid_1\phi_2\phi_3
\Big(-\mu+\alpha'(-2\nabla_3^2+\nabla_1^2)\Big)f_1(x_1,x_2,x_3)\,.
\eeq
Here it has been taken into account that $f_1(x_1,x_2,x_3)$ is symmetric
in
$x_2$ and $x_3$

In full analogy we calculate the commutators related to the part with
$f_2$.
\[
[\phid_1\phid_2\phi_3,\phid(x)\phi(x)]=
\delta^2(x_3-x)\phid_1\phid_2\phi(x)-\delta^2(x_2-x)\phid(x)\phid_1\phi(x)
-\delta^2(x_1-x)\phid(x)\phid_2\phi_3\,
\]
and
\[
[\phid_1\phid_2\phi_3,\nabla\phid(x)\nabla\phi(x)]
\]\[
=
\nabla\delta^2(x_3-x)\phid_1\phid_2\nabla\phi(x)-
\nabla\delta^2(x_2-x)\nabla\phid(x)\phid_1\phi(x)
-\nabla\delta^2(x_1-x)\nabla\phid(x)\phid_2\phi_3\,,
\]
which gives the second part of $[Q_1,H_0]$:
\beq
[Q_1^{(2)},H_0]=2\frac{i}{\mu}
\int \prod_{i=1}^3d^2x_i\phid_1\phid_2\phi_3
\Big(-\mu+\alpha'(-2\nabla_1^2+\nabla_3^2)\Big)f_2(x_1,x_2,x_3)\,.
\eeq
It has been taken into account that $f_2(x_1,x_2,x_3)$ is symmetric
in $x_1$ and $x_2$.

To satisfy the first of the conditions (\ref{condq}) which determines the
form of $Q_1$ we have to require
\beq
\Big(-\mu+\alpha'(-2\nabla_3^2+\nabla_1^2)\Big)f_1(x_1,x_2,x_3)=
-\mu\delta^2(x_1-x_2)\delta^2(x_1-x_3)
\eeq
and
\beq
\Big(-\mu+\alpha'(-2\nabla_1^2+\nabla_3^2)\Big)f_2(x_1,x_2,x_3)=
\mu\delta^2(x_3-x_1)\delta_2(x_3-x_2)\,.
\eeq

These equations are trivially solved in the momentum space.
We define the Fourier transforms by
\beq
f_i(k_1,k_2,k_3)=\int\prod_{i=1}^3\Big(d^2x_ie^{-ik_ix_i}\Big)
f_i(x_1,x_2,x_3),\ \ i=1,2 \,.
\eeq
Then we find
\beq
\Big(-\mu+\alpha'(2k_3^2-k_1^2)\Big)f_1(k_1,k_2,k_3)=
-\mu(2\pi)^2\delta^2(k_1+k_2+k_3)
\eeq
and
\beq
\Big(-\mu+\alpha'(2k_1^2-k_3^2)\Big)f_2(k_1,k_2,k_3)=
\mu(2\pi)^2\delta^2(k_1+k_2+k_3)\,.
\eeq
So taking into account the symmetry properties of the functions
$f_1$ and $f_2$ one has
\beq
f_1(k_1,k_2,k_3)=\mu\frac{(2\pi)^2\delta(k_1+k_2+k_3)}
{\mu-\alpha'(k_2^2+k_3^2-k_1^2)}
\label{f1}
\eeq
and
\beq
f_2(k_1,k_2,k_3)=-\mu\frac{(2\pi)^2\delta(k_1+k_2+k_3)}
{\mu-\alpha'(k_1^2+k_2^2-k_3^2)}=-f_1(k_3,k_2.k_1)\,.
\label{f2}
\eeq
Note that the denominators in (\ref{f1}) and (\ref{f2})
may vanish. The requirement that $Q_1$ be a Hermithean operator
implies that the singularities in (\ref{f1}) and (\ref{f2}) be
circumvented from opposite sides or taken both in the principle value 
prescription. The study of the transformed Hamiltonian in the next two sections
reveals that at order $\lambda^2$ there appears a pairwise interaction
of pomerons similar to the pairwise interaction of normal non-relativistic 
particles. In the latter case the interaction potential is standardly 
real, which leads to invariance under  time reflection. The pomerons are not 
propagating in real time, but rather in the imaginary one corresponding to
the rapidity. Still, as mentioned above there exists a similar
invariance consisting in changing signs of $\phi$ and $\phid$
and taking complex conjugate of the coefficient functions. This requires
the pair potential to be real as in the normal theory. As we shall see in 
the next section this requirement requires that the functions
$f_1$ and $f_2$ be real, so that their  singularities be taken
in the principal value sense. This circumstance will be implicitly
understood in the following.
  
\section{The transformed Hermithean Hamiltonian $h^{(2)}$}
In this section we shall find the second order Hamiltonian $h^{(2)}$
determined by second of Eqs. (\ref{hamilh}).
We present the interaction term $H_I$ as an integral over three momenta
$q_i$, $i=1,2,3$ using
\beq
\phi(x)=\int\frac{d^2k}{2\pi}e^{ikx}\phi(k)
\eeq
and similarly for $\phid(x)$. With this normalization the fields will
obey the standard commutation relations in the momentum space
\beq
[\phi(k),\phid(k')]=\delta^2(k-k')\,.
\eeq
The interaction Hamiltonian acquires the form
\[
H_I=i\int\prod_{i=1}^3\frac{d^2q_i}{2\pi}\,\Big[
(2\pi)^2\delta^2(q_1-q_2-q_3)\phid(q_1)\phi(q_2)\phi(q_3)
\]\beq
+
(2\pi)^2\delta^2(q_1+q_2-q_3)\phid(q_1)\phid(q_2)\phi(q_3)\Big]\,.
\eeq
In the same manner we get operator $Q_1$ as
\beq
Q_1=-2\frac{i}{\mu}\int\prod_{i=1}^3\frac{d^2k_i}{2\pi}\Big(f_1(-k_1,k_2,k_3)
\phid(k_1)\phi(k_2)\phi(k_3)+
f_2(-k_1,-k_2,k_3)\phid(k_1)\phid(k_2)\phi(k_3)\Big)\,.
\eeq

To find the second order Hamiltonian $h^{(2)}$ we have to calculate the
following 4 commutators:
\[
C_1=[\phid(k_1)\phi(k_2)\phi(k_3),\phid(q_1)\phi(q_2)\phi(q_3)]\,,
\]
\[
C_2=[\phid(k_1)\phi(k_2)\phi(k_3),\phid(q_1)\phid(q_2)\phi(q_3)]\,,
\]
\[
C_3=[\phid(k_1)\phid(k_2)\phi(k_3),\phid(q_1)\phi(q_2)\phi(q_3)]\,,
\]
\[
C_4=[\phid(k_1)\phid(k_2)\phi(k_3),\phid(q_1)\phid(q_2)\phi(q_3)]\,.
\]
They are all trivially found:
\[
C_1=-\delta^2(q_2-k_1)\phid(q_1)\phi(q_3)\phi(k_2)\phi(k_3)
-\delta^2(q_3-k_1)\phid(q_1)\phi(q_2)\phi(k_2)\phi(k_3)
\]\beq
+\delta^2(q_1-k_2)\phid(k_1)\phi(k_3)\phi(q_2)\phi(q_3)
+\delta^2(q_1-k_3)\phid(k_1)\phi(k_2)\phi(q_2)\phi(q_3)\,,
\eeq
\[
C_2=-\delta^2(k_1-q_3)\phid(q_1)\phid(q_2)\phi(k_2)\phi(k_3)
\]\[
+\delta(k_2-q_2)\phid(q_1)\phid(k_1)\phi(k_3)\phi(q_3)
+\delta(k_3-q_2)\phid(q_1)\phid(k_1)\phi(k_2)\phi(q_3)
\]\beq
+\delta(k_2-q_1)\phid(k_1)\phi(k_3)\phid(q_2)\phi(q_3)
+\delta(k_3-q_1)\phid(k_1)\phi(k_2)\phid(q_2)\phi(q_3)\,,
\eeq
\[
C_3=-C_2(k_1,k_2,k_3\leftrightarrow q_1,q_2,q_3)\]\[=
\delta^2(q_1-k_3)\phid(k_1)\phid(k_2)\phi(q_2)\phi(q_3)
-\delta(q_2-k_2)\phid(k_1)\phid(q_1)\phi(q_3)\phi(k_3)
\]\[
-\delta(q_3-k_2)\phid(k_1)\phid(q_1)\phi(q_2)\phi(k_3)
-\delta(q_2-k_1)\phid(q_1)\phi(q_3)\phid(k_2)\phi(k_3)
\]\beq
-\delta(q_3-k_1)\phid(q_1)\phi(q_2)\phid(k_2)\phi(k_3)\,,
\eeq
\[
C_4=C_1^{\dagger}(k_1,k_2,k_3\leftrightarrow q_3,q_2,q_1)\]\[=
-\delta^2(k_2-q_3)\phid(q_1)\phid(q_2)\phid(k_1)\phi(k_3)
-\delta^2(k_1-q_3)\phid(q_1)\phid(q_2)\phid(k_2)\phi(k_3)
\]\beq
+\delta^2(k_3-q_2)\phid(k_1)\phid(k_2)\phid(q_1)\phi(q_3)
+\delta^2(k_3-q_1)\phid(k_1)\phid(k_2)\phid(q_2)\phi(k_3)\,.
\eeq

Passing to $h^{(2)}$ we get the following 4 terms
\[
h^{(2)}_1=-\frac{1}{2}\int\prod_{i=1}^3\frac{d^2k_id^2q_i}{(2\pi)^2}
\frac{(2\pi)^2\delta^2(q_1-q_2-q_3)(2\pi)^2\delta(k_1-k_2-k_3)}
{\mu-\alpha'(k_2^2+k_3^2-k_1^2)}C_1(k_1,k_2,k_3|q_1,q_2,q_3)
\]\[
=-\frac{1}{(2\pi)^2}\int \frac{d^2k_2d^2k_3d^2q_2}
{\mu-\alpha'(k_2^2+k_3^2-(k_2+k_3)^2)}\]\beq
\Big(\phid(k_2+k_3)\phi(k_3)\phi(q_2)\phi(k_2-q_2)-
\phid(q_2+k_2+k_3)\phi(q_2)\phi(k_2)\phi(k_3)\Big)\,,
\eeq

\[
h^{(2)}_2=-\frac{1}{2}\int\prod_{i=1}^3\frac{d^2k_id^2q_i}{(2\pi)^2}
\frac{(2\pi)^2\delta^2(q_1+q_2-q_3)(2\pi)^2\delta(k_1-k_2-k_3)}
{\mu-\alpha'(k_2^2+k_3^2-k_1^2)}C_2(k_1,k_2,k_3|q_1,q_2,q_3)\]\[=
-\frac{1}{2}\frac{1}{(2\pi)^2}\int \frac{d^2k_2d^2k_3d^2q_1}
{\mu-\alpha'(k_2^2+k_3^2-(k_2+k_3)^2)}
\Big(2\phid(q_1)\phid(k_2+k_3)\phi(k_3)\phi(q_1+k_2)\]\beq+
2\phid(k_2+k_3)\phi(k_3)\phid(q_1)\phi(q_1+k_2)-
\phid(q_1)\phid(k_2+k_3-q_1)\phi(k_2)\phi(k_3)\Big)\,,
\eeq

\[
h^{(2)}_3=+\frac{1}{2}
\int\prod_{i=1}^3\frac{d^2k_id^2q_i}{(2\pi)^2}
\frac{(2\pi)^2\delta^2(q_1-q_2-q_3)(2\pi)^2\delta(k_1+k_2-k_3)}
{\mu-\alpha'(k_1^2+k_2^2-k_3^2)}C_3(k_1,k_2,k_3|q_1,q_2,q_3)\]
\[=
-\frac{1}{2}\frac{1}{(2\pi)^2}\int \frac{d^2k_1d^2k_2d^2q_3}
{\mu-\alpha'(k_1^2+k_2^2-(k_1+k_2)^2)}
\Big(2\phid(k_1)\phid(k_2+q_3)\phi(q_3)\phi(k_1+k_2)\]\beq+
2\phid(k_2+q_3)\phi(q_3)\phid(k_1)\phi(k_1+k_2)-
\phid(k_1\phid(k_2)\phi(k_1+k_2-q_3)\phi(q_3)\Big)\,,
\eeq

\[
h^{(2)}_4=+\frac{1}{2}
\int\prod_{i=1}^3\frac{d^2k_id^2q_i}{(2\pi)^2}
\frac{(2\pi)^2\delta^2(q_1+q_2-q_3)(2\pi)^2\delta(k_1+k_2-k_3)}
{\mu-\alpha'(k_1^2+k_2^2-k_3^2)}C_4(k_1,k_2,k_3|q_1,q_2,q_3)\]\[=
-\frac{1}{(2\pi)^2}\int\frac{d^2k_2d^2k_3d^2q_2}
{\mu-\alpha'(k_2^2+k_3^2-(k_2+k_3)^2)}\]\beq
\Big(\phid(k_2-q_2)\phid(q_2)\phid(k_3)\phi(k_2+k_3)-
\phid(k_3)\phid(k_2)\phid(q_2)\phi(k_2+k_3+q_2)\Big)\,.
\eeq

We have obviously
\[ h^{(2)}_4=\Big(h^{(2)}_1\Big)^\dagger,\ \
h^{(2)}_3=\Big(h^{(2)}_2\Big)^\dagger\,,\]
so that the total second order Hamiltonian $h$ is Hermithean.

\section{Single- and two-particles terms in $h$ and renormalization}
All terms in $h^{(2)}$ split into the pomeron number conserving,
$h^{(2)}_2+h^{(2)}_3$, and pomeron number changing:
$h^{(2)}_1$, with $\Delta N=-2$ and $h^{(2)}_4$ with $\Delta N=+2$.
The two latter terms will contribute to energy levels only in the
order $\lambda^4$ and can be neglected in the order $\lambda^2$
with we restrict ourselves here. 

The pomeron number conserving terms may be rewritten in the normal form.
As a result we get two contributions with two or four field
operators, describing single- and double- particle parts.
The single particle part is found to be
\[
h^{(2)}_{single}=
-\frac{2}{(2\pi)^2}{\rm Re}\,\int \frac{d^2k_2d^2k_3}
{\mu-\alpha'(k_2^2+k_3^2-(k_2+k_3)^2)}
\phid(k_2+k_3)\phi(k_2+k_3)\]\beq=
\int d^2k\phid(k)\phi(k)\Delta^{(2)}\epsilon(k)\,,
\eeq
where
\beq
\Delta^{(2)}\epsilon(k)=-
\frac{2}{(2\pi)^2} {\rm Re}\,
\int \frac{d^2k_2d^2k_3\delta^2(k_2+k_3-k)}
{\mu-\alpha'(k_2^2+k_3^2-k^2)}
\label{deleps}
\eeq
is the shift in the pomeron energy in order $\lambda^2$, so that the total
pomeron energy is
\beq
\epsilon(k)=-\mu+\alpha'k^2+\lambda^2\Delta^{(2)}\epsilon(k)\,.
\label{eps}
\eeq
One easily finds that $\Delta^{(2)}\epsilon(k)$ can be presented (see Appendix
A) as
\beq
\Delta^{(2)}\epsilon(k)=+\frac{1}{4 \pi \alpha'}\int_0^{\infty}\frac{dx}{x}
e^{x(\mu+\alpha'k^2/2)} \,.
\eeq
The integral obviously exists only for $\mu<-\alpha'k^2/2$ and diverges at $x=0$.
To regularize it we
require that at $k=0$ the change of pomeron energy vanishes,
that is $\epsilon(0)=-\mu$. In a way this is a definition of the
renormalized intercept. One can obtain the same result choosing the standard
dimensional regularization approach. With this condition the regularized
$\Delta\epsilon(k)$ is
\[
\Delta^{(2)}\epsilon_{reg}(k)=
+\frac{1}{4\pi \alpha'}\int_0^{\infty}\frac{dx}{x}
\Big(e^{x(\mu+\alpha'k^2/2)}-e^{x\mu}\Big)\]\beq=
-\frac{1}{4 \pi \alpha'}\ln \Big(1+\frac{\alpha' k^2}{2\mu}\Big)\,.
\eeq
It exists for any values of $\mu$ and is real for $\mu>0$.
Taking into account that the initial pomeron trajectory is
determined only up to terms linear in $k^2$, we approximate
the energy shift as
\beq
\Delta^{(2)}\epsilon_{reg}(k)=-\frac{1}{8\pi \mu}k^2\,,
\eeq
so that the net effect of the single-particle term in $h^{(2)}$ is to
renormalize the slope:
\beq
\alpha'\to\alpha'_{ren}=\alpha'-\lambda^2\frac{1}{8\pi \mu}\,.
\eeq

The two-pomeron interaction term $h^{(2)}_{pair}$ can be presented
in the form corresponding to transition $k_1,k_2\to q_1,q_2$
\beq
h^{(2)}_{pair}=\int d^2k_1d^2k_2d^2q_1d^2q_2\delta^2(q_1+q_2-k_1-k_2)
V^{(2)}(q_1,q_2|k_1,k_2)\phid(q_1)\phid(q_2)\phi(k_1)\phi(k_2)\,,
\eeq
where
\[
V^{(2)}(q_1,q_2|k_1,k_2)=
-\frac{1}{2\pi^2}\frac{1}{\mu-\alpha'(k_1^2+(k_2-q_1)^2-q_2^2)}
-\frac{1}{2\pi^2}\frac{1}{\mu-\alpha'(q_1^2+(q_2-k_1)^2-k_2^2)}
\]\beq
+\frac{1}{8\pi^2}\frac{1}{\mu-\alpha'(k_1^2+k_2^2-(k_1+k_2)^2)}
+\frac{1}{8\pi^2}\frac{1}{\mu-\alpha'(q_1^2+q_2^2-(q_1+q_2)^2)}
\eeq
(symmetrization in $q_1,q_2$ and $k_1,k_2$ is implied). As mentioned,
poles
in the two terms have to be understood in the principal value sense
for
the potential to be not only Hermithean but also real. 

This potential is non-local and degenerate. To more clearly see its
properties consider a  case when $q_1+q_2=k_1+k_2=0$ corresponding to the
two-pomerom exchange for the forward scattering amplitude.
Then denoting
$q_1=-q_2=q$ and $k_1=-k_2=k$ we have a potential
\[
V^{(2)}(q|k)=
-\frac{1}{2\pi^2}\frac{1}{\mu-\alpha'(k^2+(k+q)^2-q^2)}
-\frac{1}{2\pi^2}\frac{1}{\mu-\alpha'(q^2+(k+q)^2-k^2)}
\]\beq
+\frac{1}{8\pi^2}\frac{1}{\mu-2\alpha'k^2}
+\frac{1}{8\pi^2}\frac{1}{\mu-2\alpha'q^2}\,.
\label{v0}
\eeq
The last two terms  depend only on the initial or only on the final 
momenta.
Since  the potential falls rather slowly at
high momenta, its integration over $q$ or $k$ meets with a logarithmic
divergence. (See the form of the kernel in the coordinate space in Appendix 2.)

The effect of this degenerate pair potential is not quite clear in the general case.
Considerations in the next section, restricted to the case
$q_1+q_2=k_1+k_2=0$, tell that the spectrum of the pomeron states will
not be changed by this interaction. So in its presence the two-pomeron
states will continue to have their total energy
\beq
E_2(k_1,k_2)=\epsilon(k_1)+\epsilon(k_2)\,,
\eeq
with $\epsilon(k)$ given by (\ref{eps}) but the wave functions will
become changed by the standard scattering operator. To the second
order in $\lambda$
\beq
\Psi_{k_1,k_2}(q_1,q_2)={\rm
Sym}\,\Big\{\delta^2(k_1-q_1)\delta^2(k_2-q_2)+
\lambda^2\frac{V(q_1,q_2|k_1,k_2)}{\epsilon(q_1)+\epsilon(q_2)-
\epsilon(k_1)-\epsilon(k_2)\pm i0}\Big\}\,,
\eeq
where symbol ${\rm Sym}$ means symmetrization in $q_1$ and $q_2$
and signs of $i0$ correspond to in- or -outgoing waves.

For the asymptotic of the Green function we shall find from the
two-pomeron states
\beq
\delta^2(k-k')\int d^2k_1d^2k_2\langle 0|\phi(k')e^{Q/2}|\Psi_{k_1,k_2}\rangle
e^{-yE_2(k_1,k_2)}\langle\Psi_{k_1,k_2}|e^{-Q/2}\phid(k)|0\rangle\,,
\eeq
where the matrix elements can be easily computed by perturbations
in $\lambda$. This asymptotics will be true at large $y$ until
$y\sim 1/\lambda^4$.
Of course one also will have similar contributions from states with the
number of pomerons greater than two, but the corresponding matrix
elements will be of the higher order in $\lambda$.

\section{ The Schroedinger equation with a degenerate potential}
\subsection{Problem}
Consider the Schroedinger equation in the 2-dimensional momentum space
\beq
(\epsilon(q)-E)\psi(q)=-\int d^2kV(q|k)\psi(k)\,.
\label{eq1}
\eeq
To simplify notation we rescale $E$ to exclude all terms independent of 
$q$ in $\epsilon(q)$
and have in our case $\epsilon(q)=2\alpha'q^2$. Our  potential has a 
structure
(\ref{v0}):
\beq
V(q,k)=v(q)+v(k)+V_1(q,k),
\label{eq2}
\eeq
where $V_1(q|k)$ has the normal properties and vanishes as any of the
arguments go to infinity. Our aim is to study the spectrum $E$ of the
solutions to Eq. (\ref{eq1})

Obviously Eq. (\ref{eq1}) can have both solutions corresponding to the
scattering states and to bound states. In the former case we
standardly
convert this equation into the Lippman-Schwinger equation presenting
\beq
\psi_l(q)=\delta^2(q-l)+\frac{T(q|l)}{\epsilon(l)-\epsilon(q)\pm i0},
\eeq
where $T(q|l)$ ($T$ from now is no more the ``time reflection'' operator) satisfies
\beq
T(q|l)=V(q|l)+\int d^2k\frac{V(q,k)T(k|l)}{\epsilon(l)-\epsilon(k)\pm
    i0}.
\label{eq3}
\eeq
If this equation can be solved it corresponds to the scattering state
with the incident momentum $l$ and energy $E=\epsilon(l)$ which
belongs to the continuous positive spectrum.

For the bound state $\psi_E(q)$ with energy $E<0$ we analogously present
\beq
\psi_E(q)=\frac{t_E(q)}{E-\epsilon(q)},
\eeq
with an equation for $t_E$
\beq
t_E(q)=\int d^2k\frac{V(q|k)t_E(k)}{E-\epsilon(k)}.
\label{eq4}
\eeq
Our aim is to study possible solutions of Eqs. (\ref{eq3}) and
(\ref{eq4})
with a degenerate potential (\ref{eq2}).

\subsection{Continuous spectrum}
Putting (\ref{eq2}) into (\ref{eq3}) and
suppressing the fixed argument $l$ in $T$ we have
\[
T(q)=v(q)+c+V_{1}(q,l)+v(q)
\int d^2k\frac{T(k)}{\epsilon(l)-\epsilon(k)}\]
\beq
+\int d^2k\frac{v(k)T(k)}{\epsilon(l)-\epsilon(k)}+
\int d^2k\frac{V_1(q,k)T(k)}{\epsilon(l)-\epsilon(k)}.
\eeq
where we have denoted the part of $V$ independent of $q$
\beq v(l)=c. \eeq
We also denote
\beq
d=\int d^2k\frac{T(k)}{\epsilon(l)-\epsilon(k)},\ \
e=\int d^2k\frac{v(k)T(k)}{\epsilon(l)-\epsilon(k)}.
\eeq
We find an equation
\beq
T(q)=c+e+(1+d)v(q)+V_1(q,l)+\int
d^2k\frac{V_1(q,k)T(k)}{\epsilon(l)-\epsilon(k)}.
\label{eqt}
\eeq

Correspondingly we present
\beq
T(q)=c+e+(1+d)v(q)+T_1(q)\,.
\label{tviat1}
\eeq
The equation for $T_1(q)$ is
\beq
T_1(q)=(c+e)\chi_1(q)+(1+d)\chi_2(q)+V_1(q,l)+
\int d^2k\frac{V_1(q,k)T_1(k)}{\epsilon(l)-\epsilon(k)},
\label{eqt1}
\eeq
with
\beq
\chi_1(q)=\int d^2k\frac{V_1(q,k)}{\epsilon(l)-\epsilon(k)}
\eeq
and
\beq
\chi_2(q)=\int d^2k\frac{V_1(q,k)v(k)}{\epsilon(l)-\epsilon(k)}.
\eeq

For the two constants $d$ and $e$ we obtain equations following
from their definition
\beq
d=(1+d)I_1+(c+e)I_0+
\int d^2k\frac{T_1(k)}
{\epsilon(l)-\epsilon(k)}
\label{eqd}
\eeq
and
\beq
e=(c+e)I_1+(1+d)I_2+
\int d^2k\frac{v(k)T_1(k)}
{\epsilon(l)-\epsilon(k)},
\label{eqe}
\eeq
where
\beq
I_n=\int d^2k \frac{v^n(k)}{\epsilon(l)-\epsilon(k)}.
\label{eqint}
\eeq

To solve Eq. (\ref{eqt1}) with additional conditions
(\ref{eqd}) and (\ref{eqe}) we first solve this equation for three
different inhomogeneous terms
\beq
T_1^{(i)}(q)=T_0^{(i)}(q)+
\int d^2k\frac{V_1(q,k)T_1^{(i)}(k)}{\epsilon(l)-\epsilon(k)}
\label{eqt1i}\,,
\eeq
where $i=1,2,3$ and
\beq
T_0^{(1,2)}(q)=\chi_{1,2}(q),\ \ T_0^{(3)}(q)=V_1(q,l)\,.
\eeq
From these three solutions we obtain the solution to
Eq. (\ref{eqt1}) as
\beq
T_1(q)=(c+e)T_1^{(1)}(q)+(1+d)T_1^{(2)}(q)+T_1^{(3)}(q)
\label{solt1}
\eeq

Now we put this solution into the equations 
(\ref{eqd}), (\ref{eqe}) to obtain a system of two linear equations
for $d$ and $e$:
\[
d(I_1+J_2-1)+e(I_0+J_1)+c(I_0+J_1)+I_1+J_2+J_3=0\,,
\]
\beq
d(I_2+K_2)+e(I_1+K_1-1)+c(I_1+K_1)+I_2+K_2+K_3=0\,,
\label{eqde}
\eeq
where $I_n$ are defined by (\ref{eqint}) and
\beq
J_n=\int d^2k \frac{T_1^{(n)}(k)}{\epsilon(l)-\epsilon(k)},\ \ 
K_n=\int d^2k \frac{v(k)T_1^{(n)}(k)}{\epsilon(l)-\epsilon(k)}.
\label{eqjk}
\eeq
All quantities in fact depend on the fixed momentum $l$.
Solution of the linear system (\ref{eqde}) is of course trivial.
The determinant is
\beq
D=(I_1+J_2-1)(I_1+K_1-1)-(I_0+J_1)(I_2+K_2)
\eeq
and so
\beq
d=\frac{1}{D}\Big[\Big(c(I_1+K_1)+I_2+K_2+K_3\Big)(I_0+I_1)-
\Big(c(I_0+J_1)+I_1+J_2+J_3\Big)(I_1+K_1-1)\Big]\,,
\label{sold}
\eeq
\beq
e=\frac{1}{D}\Big[\Big(c(I_0+J_1)+I_1+J_2+J_3\Big)(I_2+K_2)-
\Big(c(I_1+K_1)+I_2+K_2+K_3\Big)(I_1+J_2-1)\Big]\,.
\label{sole}
\eeq
Of course the resulting $d$ and $e$ are functions of the fixed momentum 
$l$.
With thus determined $d(l)$ and $e(l)$ Eq. (\ref{solt1}) gives the final 
solution to the Lippmann-Schwinger problem. The only difficulty
may generally arise in case $D(l)=0$ which may only happen at some 
specific values 
of $l$ and leads to certain singularities of the scattering matrix at 
these values of momentum, which we consider improbable.

For future reference, here we present  orders in $\lambda$ for 
different quantities defined in the previous derivation.
Obviously 
\[
D=1+{\cal O}(\lambda^2);\ \ I_n\sim \lambda^{2n};
\ \ \chi_{1(2)}\sim \lambda^{2(4)}\]\beq 
T_1^{(1)},\ T_1^{(3)},\ J_1,\ J_3\sim\lambda^2;\ \ 
T_1^{(2)},\ J_2\sim\lambda^4;\ \ 
K_1,K_3\sim\lambda^4;\ \ K_2\sim\lambda^6\,.
\label{order1}
\eeq 
Then it follows from (\ref{sold}) and (\ref{sole}) that
$d\sim \lambda^2$ and $e\sim\lambda^4$ and in the lowest
approximation (order $\lambda^2$) the scattering matrix $T$ is given
just by the total potential $V$, as expected.

However in our case there is a new problem. The constant $I_0$
is in fact divergent. In the limit $I_0\to\infty$ we find that the
determinant grows linearly with $I_0$:
\beq
D=-I_0(I_2+K_2)\,.
\label{dlim}
\eeq
The denominators of (\ref{sold}) and (\ref{sole}) also grow linearly
with $I_0$. So in the limit $I_0\to\infty$ we find finite values for both
$d$ and $e$:
\beq
1+d=-\frac{c+K_3}{I_2+K_2},\ \ e+c=0\,.
\label{soldelim}
\eeq
With these values we find in this limiting case 
\beq
T_1(q)=T_1^{(3)}(q)-\frac{c+K_3}{I_2+K_2}T_1^{(2)}(q)\,.
\label{solt1lim}
\eeq
So again the solution in all probability exists but the constant $e$
is automatically adjusted to exclude the constant term $c=v(l)$
from the original Lippmann-Schwinger equation
(\ref{eqt}). 

As a result, we find that even for divergent $I_0$
the Lippmann-Schwinger equation has a solution for any positive
energy, so that the spectrum is continuous and covers all positive
values of energy

Note that in the limit $I_0\to\infty$ orders of $d$ and $e$
in powers of $\lambda$ are radically changed. Now the determinant
$D\sim \lambda^4$ and as a result $1+d\sim 1/\lambda^2$ and
$e\sim \lambda^2$. As a result already in the lowest order $\lambda^2$
the scattering matrix $T_1$ acquires additional terms from $T_1^{(2)}$: 
\beq
T_1(q)=V_1(q|l)-\frac{v(l)}{I_2}\chi_2(q)\,.
\label{limt1}
\eeq
Turning to the full scattering matrix (\ref{tviat1}) we find that it
acquires a term of the order unity
\beq
T(q|l)=-\frac{v(q)v(l)}{I_2}+T_1(q)\,.
\label{limt}
\eeq
It may be considered as a renormalization term for the scattering
matrix in the limit $I_0\to\infty$. Of course appearance of this term
is due to the implicitly made assumption that
$\lambda^4I_0>>1$ as $I_0\to\infty$. Different relations between the 
small $\lambda$ and large $I_0$ will lead to different results.
  
To illustrate the described procedure for the solution of 
Lippmann-Schwinger equation in Appendix C we calculate the scattering
matrices $T$ and $T_1$ up to order $\lambda^2$ for the pair pomeron 
potential $V^{(2)}$, Eq.(\ref{v0}), for the forward case 
$q_1+q_2=k_1+k_2=0$. The found expressions are long and not very 
interesting but they show that the procedure is quite feasible and does 
not
encounter any new complications.

 \subsection{Discrete spectrum}
Putting (\ref{eq2}) into (\ref{eq4})
we now obtain
\beq
t_E(q)=v(q)
\int d^2k\frac{t_E(k)}{E-\epsilon(k)}
+\int d^2k\frac{v(k)t_E(k)}{E-\epsilon(k)}+
\int d^2k\frac{V_1(q,k)t_E(k)}{E-\epsilon(k)}.
\eeq
As before
we  denote
\beq
d=\int d^2k\frac{t_E(k)}{E-\epsilon(k)},\ \
e=\int d^2k\frac{v(k)t_E(k)}{E-\epsilon(k)}.
\eeq
We find an equation
\beq
t_E(q)=e+dv(q)+\int
d^2k\frac{V_1(q,k)t_E(k)}{E-\epsilon(k)}.
\eeq

We present
\beq
t_E(q)=e+dv(q)+t_{1E}(q)
\eeq
to find an equation for $t_{1E}$
\beq
t_{1E}(q)=e\chi_1(q)+d\chi_2(q)+\int d^2k\frac{V_1(q,k)t_{1E}(k)}
{E-\epsilon(k)},
\label{eqtb}
\eeq
where similarly to the continuous spectrum case
\beq
\chi_1(q)=\int d^2k\frac{V_1(q,k)}{E-\epsilon(k)}
\eeq
and
\beq
\chi_2(q)=\int d^2k\frac{V_1(q,k)v(k)}{E-\epsilon(k)},
\eeq
with the two conditions to determine $d$ and $e$
\beq
d=dI_1+eI_0+\int d^2k\frac{t_{1E}(k)}{E-\epsilon(k)}
\label{eqdb}
\eeq
and
\beq
e=eI_1+dI_2+\int d^2k\frac{v(k)t_{1E}(k)}{E-\epsilon(k)}.
\label{eqeb}
\eeq
The integrals $I_n$ are the same as in (\ref{eqint}) with
$\epsilon(l)\to E$.

Obviously the solution to Eq. (\ref{eqtb}) can be presented as a sum
\beq
t_{1E}(q)=e\,t_{1E}^{(1)}(q)+d\,t_{1E}^{(2)}(q),
\eeq
where the two functions $t_{1E}^{(1,2)}$ satisfy
\beq
t_{1E}^{(i)}(q)=\chi_1(q)+\int d^2k\frac{V_1(q,k)t_{1E}^{(i)}(k)}
{E-\epsilon(k)},\ \ i=1,2
\eeq

After functions  $t_{1E}^{(1,2)}$ are known, one finds a homogeneous
system of linear equations to determine $d$ and $e$:
\beq
d=eI_0+dI_1+eJ_1+dJ_2,\ \
e=eI_1+dI_2+eK_1+dK_2,
\label{syseq}
\eeq
where now, similarly to (\ref{eqjk}),
\beq
J_n=\int d^2k \frac{t_{1E}^{(n)}(k)}{E-\epsilon(k)},\ \ 
K_n=\int d^2k \frac{v(k)t_{1E}^{(n)}(k)}{E-\epsilon(k)}.
\label{eqjk1}
\eeq
All the coefficients in the system (\ref{syseq}) depend on $E$.
The value of the bound energy $E$ is found from the condition of
existence of solutions to the system (\ref{syseq}):
\beq
(I_1+J_2-1)(I_1+K_1-1)-(I_0+J_1)(I_2+K_2)=0.
\label{ebound}
\eeq

Now consider the case $I_0\to\infty$. Then Eq. (\ref{ebound})
reduces to
\beq
I_2(E)+K_2(E)=0,
\label{ebound1}
\eeq
which determines a possible bound state in this limiting case.

As we have seen, at small values of $\lambda$
$I_2$ is of  order $\lambda^4$ and 
$K_2$ is of order $\Lambda^6$
Since $I_2(E)$ cannot
vanish (it is strictly negative for $E<0$) Eq.(\ref{ebound1}) cannot
be satisfied. So for small values of $\lambda$ there are no bound states
when $I_0\to\infty$.
\section{Conclusions}
We have generalized the technique of constructing an Hermithean
Hamiltonian for the PT symmetric  LRFT model, developed in
~\cite{BV} for the toy model in zero transverse dimensions,
to the realistic case of two transverse dimensions.
The complexity of the latter model makes both the derivation and
analysis of the found Hamiltonian not straightforward already in the
lowest non-trivial order in the coupling constant $\lambda$ of the triple
pomeron interaction. In particular the divergence of the pomeron
intercept has to be eliminated by renormalization. Also the found 
pair interaction between pomerons is both singular and degenerate.
It has required a separate study  of the Schroedinger equation
with degenerate potentials, which may have a wider scope of
applicability. 

As a result we have found that at small $\lambda$
the total impact of the pomeron interaction at order $\lambda^2$ 
is reduced to the change of slope. 
These results allow to study the asymptotic of any scattering amplitude
at large rapidities $y$  in the region 
\[1<<y<<(\alpha'/\lambda^2)^2\]
However in the course of our study 
it became clear that at finite $\lambda$ the pomerons may form
bound states, whose presence will drastically change this asymptotics.

Note that the experimental values for $\alpha'$ and $\lambda$
are roughly $\alpha'\sim 0.25$ (GeV/c)$^{-2}$ and $\lambda\sim0.33$ 
(GeV/c)$^{-1}$, so that the actual parameter of the perturbative expansion
is $\lambda^2/\alpha' \sim 0.4$, which is not so small. So to estimate
the possibility to apply our results to the realistic processes one
has to study the NNLO (terms of order $\lambda^4$). If they happen to be 
relatively small then one can study possible pomeron bound states
using our potential and, say, the variational methods.

We stress that the perturbative study of our Hermithean Hamiltonian
cannot give the true asymptotic of the theory at $y\to\infty$, which
remains unperturbative. To find it one has to search for non-perturbative
techniques to construct this Hamiltonian.

\section{Akcnowledgments}
M.A.B. greatly acknowledges hospitality and financial support of
INFN, Sezione Bologna, where this paper was completed.
  
\appendix
\section{Single pomeron energy shift}
We have to calculate the integral
\beq
I=\int \frac{d^2k_2d^2k_3\delta^2(k_2+k_3-k)}
{\mu-\alpha'(k_2^2+k_3^2-k^2)}\,.
\eeq
We present the $\delta$-function as an integral over ${\bf r}$
and the denominator at $\mu<-\alpha'k^2/2$ as
\beq
\frac{1}{\mu-\alpha'(k_2^2+k_3^2-k)^2}=-\int_0^{\infty}dx
e^{x(\mu-\alpha'(k_2^2+k_3^2-k^2))}
\eeq
to get
\beq
I=-\frac{1}{(2\pi)^2}\int_0^{\infty}dx\int d^2k_2d^2k_3d^2re^{ir(k_2+k_3-k)}
e^{x(\mu-\alpha'(k_2^2+k_3^2-k^2))}\,.
\eeq
Integrals over $k_2$ and $k_3$ give the same result:
\beq
\int
d^2k_2e^{irk_2-x\alpha'k_2^2}=
\frac{\pi}{x\alpha'}e^{-\frac{r^2}{4x\alpha'}}\,.
\eeq
So we find
\beq
I=-\frac{1}{4}\int_0^{\infty}dx\int d^2r
\frac{1}{(x\alpha')^2}e^{-irk)}
e^{x(\mu+\alpha'k^2-\frac{r^2}{2x\alpha'})}\,.
\eeq
Next we integrate over $r$
\beq
\int d^2re^{ikr-\frac{r^2}{2x\alpha'}}=2\pi x\alpha'
e^{-x\alpha'k^2/2}
\eeq
to finally find an integral over $x$
\beq
I=-\frac{\pi}{2\alpha'}\int_0^{\infty}\frac{dx}{x}
e^{x(\mu+\alpha'k^2/2)}\,.
\eeq
\section{The coordinate space potential between pomerons}
The two-pomeron potential is non-local in the coordinate space.
We shall limit ourselves with the case $q_1+q_2=k_1+k_2$ when the in the momentum
space the potential is given by (\ref{v0}). Then in the coordinate space
its kernel is  defined as
\beq
V^{(2)}(y|x)=\int\frac{d^2qd^2k}{(2\pi)^4}e^{-iqy+ikx}V^{(2)}(q|k)\,.
\eeq
The non-degenerate part of the kernel comes from the first two terms
in (\ref{v0}) and is
\beq
V^{(2)}_1(y|x)=-
\frac{1}{2\pi^2}\int\frac{d^2qd^2k}{(2\pi)^4}e^{-iqy+ikx}
\frac{1}{\mu-\alpha'(k^2+(k+q)^2-q^2)} +\Big(x\leftrightarrow  -y\Big)\,.
\eeq
We rewrite the first term in this expression as
\[
V^{(2)}_{11}(y|x)=-
\frac{1}{2\pi^2}\int\frac{d^2qd^2kd^2\kappa}{(2\pi)^4}
\delta^2(k+q-\kappa)e^{-iqy+ikx}
\frac{1}{\mu-\alpha'(k^2+\kappa^2-q^2)}\]\beq
=
-\frac{2}{(2\pi)^8\alpha'}\int d^2qd^2kd^2\kappa d^2r
e^{-iqy+ikx+ir(q+k-\kappa)}
\frac{1}{m+ak^2+bq^2+c\kappa^2}\,.
\eeq
where $a=c=-1, b=1$ and $m=\mu/\alpha'$
We further present
\beq
\frac{1}{m+ak^2+bq^2+c\kappa^2}=\int_0^{\infty}d\xi e^{-\xi(m+ak^2+bq^2+c\kappa^2)}
\eeq
and do the Gaussian integrals in $k,q$ and $\kappa$ assuming that they exist, that is
for positive $a,b$ and $c$. Transition to the desired values of $a,b$ and $c$ will be
achieved by analytic continuation. We get
\beq
V^{(2)}_{11}(y|x)=-\frac{2}{(2\pi)^8\alpha'}
\frac{\pi^3}{abc}\int_0^{\infty}\frac{d\xi}{\xi^3}
e^{-m\xi}\int d^2re^{-(x+r)^2/4a\xi-(y-r)^2/4b\xi-r^2/4c\xi}\,.
\eeq
The exponent in the integrand in $r$ has the form
\[-(\beta r^2-2sr+x^2/a+y^2/b)/4\xi\,, \]
where
\[ \beta=1/a+1/b+1/c,\ \ s=(-x/a+y/b)\,,\]
so that after the integration over $r$ we find
\beq
V^{(2)}_{11}(y|x)=\frac{1}{32\pi^4\alpha'}
\frac{1}{abc\beta}\int_0^{\infty}\frac{d\xi}{\xi^2}
e^{-m\xi-t/\xi}\,,
\eeq
where
\beq
t=\frac{1}{4}\Big(-\frac{s^2}{\beta}+\frac{x^2}{a}+\frac{y^2}{b}\Big)\,.
\eeq
Integration over $\xi$ gives
\beq
V^{(2)}_{11}(y|x)=-\frac{1}{8\pi^4\alpha'}
\frac{1}{abc\beta}\frac{m}{z_1}{\rm K}_1(z_1)
\eeq
where $z_1=2\sqrt{mt}$
Inserting the desired values  $a=-1$, $b=1$ and  $c=-1$ we find 
$abc\beta=-1$ so that
\beq
z_1=\sqrt{\frac{2\mu{\bf y}({\bf y}+{\bf x})}{\alpha'}}
\eeq
and thus
\beq
V^{(2)}_{11}(y|x)=\frac{1}{8\pi^4}\frac{\mu}{{\alpha'}^2}\frac{1}{z_1}{\rm 
K}_1(z_1)
\label{v01}
\eeq
If $z_1$ is real then the  potential is  falls exponentially.
If $z_1$ is pure imaginary, then putting $z=-i|z|$ and taking the real 
part according to the principal value prescription the potential is 
expressed via the oscillating Neumann function:
\beq
V^{(2)}_{11}(y|x)=\frac{1}{16\pi^3}\frac{\mu}{{\alpha'}^2}
\frac{1}{|z_1|}{\rm N}_1(|z_1|)
\label{v011}
\eeq

The second part of the potential $V_1$ is obtained as
\beq
V^{(2)}_{12}(y|x)=V^{(2)}_{11}(-x|-y)=
\frac{1}{8\pi^4}\frac{\mu}{{\alpha'}^2}\frac{1}{z_2}{\rm 
K}_1(z_2)\,,
\label{v02}
\eeq
where
\beq
z_2=\sqrt{\frac{2\mu{\bf x}({\bf x}+{\bf y})}{\alpha'}}\,.
\eeq

The degenerate part of the potential is simpler.
We have
\beq
V_2^{(2)}(y|x)=\frac{1}{8\pi^2}
\int\frac{d^2qd^2k}{(2\pi)^4}e^{-iqy+ikx}
\frac{1}{\mu-2\alpha'k^2} +\Big(x\leftrightarrow  -y\Big)\,.
\eeq
We calculate the first part by taking $\mu=-2\alpha'\nu$ and
assuming $\nu>0$ to subsequently continue to negative $\nu$.
So
\beq
V_{21}^{(2)}(y|x)=-\frac{1}{4\alpha'(2\pi)^4}\delta^2(y)
\int d^2ke^{ikx}\frac{1}{k^2+\nu}\,.
\eeq
The integral in $k$ is trivially done to give
\beq
V_{21}^{(2)}(y|x)=-\frac{1}{4\alpha'(2\pi)^3}\delta^2(y)
{\rm K}_0(\sqrt{\nu x^2})\,.
\eeq
Continuing to negative $\nu$ and putting
$\sqrt{\nu x^2}=-i\sqrt{|\nu| x^2}$ we finally find
\beq
V_{21}^{(2)}(y|x)=+\frac{1}{16\alpha'(2\pi)^2}\delta^2(y)
{\rm N}_0\left(\sqrt{\frac{\mu x^2}{2\alpha'}}\right)\,.
\eeq
Taking into account the requirement of hermiticity, the second
part of the degenerate potential will be given
by
\beq
V_{22}^{(2)}(y|x)=\frac{1}{16\alpha'(2\pi)^2}\delta^2(x)
{\rm N}_0\left(\sqrt{\frac{\mu y^2}{2\alpha'}}\right)\,.
\eeq

\section{Lowest order scattering matrix for the pomeron interaction}
As in Section 4 we restrict ourself to the forward case $q_1+q_2=
k_1+k_2$ where the full potential $V(q,k)$ is given by
(\ref{v0}), so that
$V_1(q|k)$ is given by the first two terms in (\ref{v0})
and 
\beq
v(k)=\frac{1}{8\pi^2}\frac{1}{\mu-2\alpha'k^2}\,.
\eeq
Obviously in our case $I_0$ is divergent. We assume that it is regularized
in some manner (say restricing integration by $k<\Lambda$). As 
$\Lambda\to\infty$ $I_0\to\infty$ logarithmically. This leads us to the
expressions (\ref{limt1}) and (\ref{limt}) for $T_1$ and $T$ 
respectively valid up to order $\lambda^2$. To  calculate these scattering
matrices all we need is to find $\chi_2(q)$ and $I_2$, keeping in mind the
reality of the potential and the prescription given in Eq. (\ref{eq3}).

Calculation of $I_2$ is of course trivial. We find
\beq
I_2(l)=-\frac{\lambda^4}{512\pi^3{\alpha'}^3}\frac{1}{l^2-m^2}
\Big[\frac{1}{l^2-m^2}\Big(\ln\frac{m^2}{l^2}-i\pi\Big)+\frac{1}{m^2}\Big]\,,
\label{i0}
\eeq
where we use a convenient notation
\beq
m^2=\frac{\mu}{2\alpha'}\,.
\label{em}
\eeq

Calculation of $\chi_2(q|l)$ is a bit more complicated.
It consists of two terms coming from the two terms in the
potential $V_1$.
The first term can be written as
\[
\chi_2^{(1)}(q|l)=\frac{\lambda^4}{128\pi^4{\alpha'}^3}
\int\frac{d^2k}{[(k+q/2)^2-p^2](k^2-m^2)(k^2-l^2)}\]\beq=
\frac{\lambda^4}{128\pi^4{\alpha'}^3}\frac{1}{l^2-m^2}
\Big(A(l^2)-A(m^2)\Big)\,,
\label{chi21}
\eeq
where
\beq
p^2=m^2+\frac{1}{4}q^2
\eeq
and
\beq
A(l^2)=\int\frac{d^2k}{[(k+q/2)^2-p^2](k^2-l^2)}\,.
\eeq
This latter integral is conveniently calculated using the
Feynman parametrization to finally give
\beq
{\rm Re}\,A(l^2)=\frac{\pi}{\sqrt{\Delta}}
\ln\frac{(l^2-m^2-q^2/2-\sqrt{\Delta})(l^2-m^2+\sqrt{\Delta})}
{(l^2-m^2-q^2/2+\sqrt{\Delta})(l^2-m^2-\sqrt{\Delta})}\,,
\label{al1}
\eeq
where
\beq
\Delta=(l^2-m^2)^2-q^2l^2
\eeq
and in (\ref{al1}) it is assumed that $\Delta>0$. For $\Delta<0$ we find
\beq
{\rm 
Re}\,A(l^2)=\frac{2\pi}
{\sqrt{-\Delta}}\Big[{\rm arctg}\frac{l^2-m^2-q^2/2}
{\sqrt{-\Delta}}-{\rm arctg}\frac{l^2-m^2}
{\sqrt{-\Delta}}\Big]\,.
\label{al2}
\eeq
The imaginary part is
\beq
{\rm Im}\,A(l^2)=-\frac{\pi^2}{\sqrt{\Delta}}\theta(\Delta)\,.
\eeq

The expression for $A(m^2)$ is simpler due to cancellations between $p^2$ 
and $m^2$. It is real:
\beq
A(m^2)=-\frac{2\pi}{qm}{\rm arctg}\frac{q}{2m}\,.
\label{am}
\eeq

The second term in $\chi$ can be written in a manner similar to
(\ref{chi21}):
\beq
\chi_2^{(2)}(q|l)=\frac{\lambda^4}{128\pi^4{\alpha'}^3}
\int\frac{d^2k}{({\bf qk}+q^2-m^2)(k^2-m^2)(k^2-l^2)}=
\frac{\lambda^4}{128\pi^4{\alpha'}^3}\frac{1}{l^2-m^2}
\Big(B(l^2)-B(m^2)\Big)\,,
\label{chi22}
\eeq
where now
\beq
B(l^2)=\int\frac{d^2k}{({\bf kq}+q^2-m^2)(k^2-l^2)}\,.
\eeq
Integration over the azimuthal angle gives
\beq
B(l^2)=2\pi\int_0^{k_m}\frac{kdk}{(k^2-l^2)\sqrt{\Delta_1}}\,,
\label{bl1}
\eeq
where 
\beq
\Delta_1=(q^2-m^2)^2-q^2k^2
\eeq
and $k_m$ is defined
by the condition that $\Delta_1>0$. Here we have used the principal
value prescription which tells that at $\Delta<0$ the real part of
the azimuthal integral is zero.
Subsequent integration over $k$ gives
\beq
B(l^2)={\rm Re}\,\frac{\pi}{\sqrt{\Delta_2}}
\ln\frac{\sqrt{\Delta_2}+|q^2-m^2|}{\sqrt{\Delta_2}-|q^2-m^2|}
-i\frac{\pi^2}{\sqrt{\Delta_2}}\theta(\Delta_2)\,,
\label{bl2}
\eeq
where
\beq
\Delta_2=(q^2-m^2)^2-q^2l^2
\eeq
and $\Delta_2>0$. For $\Delta_2<0$ $B(l^2)$ is real and given by
\beq
B(l^2)=-\frac{2\pi}{\sqrt{-\Delta_2}}{\rm arctg}\frac{|q^2-m^2|}{\sqrt{-\Delta_2}}\,.
\label{bl3}
\eeq

Calculation of $B(m^2)$ obviously gives the same expressions (\ref{bl2}) and (\ref{bl3})
in which $l^2$ is substituted by $m^2$.
This finishes calculation of $\chi_2(q|l)$.



\begin{thebibliography}{99}
\bibitem{BFKL} E. A. Kuraev, L. N. Lipatov and V. S. Fadin,
               Sov. {\bf JETP 44} (1976) 443; \\
               {\bf ibid. 45} (1977) 199;\\
               Ya. Ya. Balitskii and L.N. Lipatov, Sov. J. Nucl. Phys.
               {\bf 28}, (1978) 822. 
%
\bibitem{BW}
  J.~Bartels,
  Z.\ Phys.\  C {\bf 60} (1993) 471;\\
  J.~Bartels and M.~Wusthoff,
  Z.\ Phys.\  C {\bf 66} (1995) 157.
%
\bibitem{Mueller}
  A.~H.~Mueller,
  Nucl.\ Phys.\  B {\bf 437} (1995) 107
  [arXiv:hep-ph/9408245];\\
 A.~H.~Mueller and B.~Patel,
  Nucl.\ Phys.\  B {\bf 425} (1994) 471
  [arXiv:hep-ph/9403256].
%
\bibitem{BV}
  M.~A.~Braun and G.~P.~Vacca,
  Eur.\ Phys.\ J.\  C {\bf 6} (1999) 147
  [arXiv:hep-ph/9711486].
%
\bibitem{Bal}
  I.~Balitsky,
  Nucl.\ Phys.\  B {\bf 463} (1996) 99
  [arXiv:hep-ph/9509348].
%
\bibitem{Kov}
  Y.~V.~Kovchegov,
  Phys.\ Rev.\  D {\bf 60} (1999) 034008
  [arXiv:hep-ph/9901281];

  Phys.\ Rev.\  D {\bf 61} (2000) 074018
  [arXiv:hep-ph/9905214].
%
%
\bibitem{B1} M.A.Braun, Phys. Lett. {\bf B 483} (2000) 115; Eur. Phys. J.
{\bf c 33} (2004) 113.
%
\bibitem{B2} M.A.Braun, Eur. Phys. J. {\bf C 48} (2006) 511.
%
\bibitem{CGS} A.H.Mueller, A.I.Shoshi and S.M.H.Wang, Nucl.Phys {\bf B 715} (2005)
440;\\
E.Iancu and D.N.Triantafyllopoulos, Phys. Lett. {\bf B 610} (2005) 253;\\
E.Levin and M.Lublinski, Nucl. Phys. {\bf A 763} (2005) 172;
S.Bondarenko, Nucl. Phys. {\bf A 792} (2007) 264.
%
%
\bibitem{schwimmer} A.Schwimmer, Nucl. Phys. {\bf B 94} (1975) 445.
%
\bibitem{acj} D.Amati, L.Caneshi and R.Jengo, Nucl. Phys. {\bf B 101}
(1975) 397.
%
\bibitem{KL}
  M.~Kozlov and E.~Levin,
  Nucl.\ Phys.\  A {\bf 779} (2006) 142
  [arXiv:hep-ph/0604039].
%
\bibitem{Bond}
  S.~Bondarenko, L.~Motyka, A.~H.~Mueller, A.~I.~Shoshi and B.~W.~Xiao,
  Eur.\ Phys.\ J.\  C {\bf 50} (2007) 593
  [arXiv:hep-ph/0609213].
%
\bibitem{BVtoy} 
  M.~A.~Braun and G.~P.~Vacca,
  Eur.\ Phys.\ J.\  C {\bf 50} (2007) 857
  [arXiv:hep-ph/0612162].
\bibitem{bender}
 C.~M.~Bender and S.~Boettcher,
  Phys.\ Rev.\ Lett.\  {\bf 80}, 5243 (1998);\\
 C.~M.~Bender, D.~C.~Brody and H.~F.~Jones,
  Phys.\ Rev.\ Lett.\  {\bf 89}, 270401 (2002)
  [Erratum-ibid.\  {\bf 92}, 119902 (2004)].
%
\bibitem{mostafazadeh}
  A.~Mostafazadeh,
  J.\ Math.\ Phys.\  {\bf 43}, 3944 (2002);\\
  A.~Mostafazadeh,
  J.\ Phys.\ A {\bf 38}, 6557 (2005)
  [Erratum-ibid.\ A {\bf 38}, 8185 (2005)].

\bibitem{BenderQFT}
  C.~M.~Bender, D.~C.~Brody and H.~F.~Jones,
  Phys.\ Rev.\  D {\bf 70} (2004) 025001
  [Erratum-ibid.\  D {\bf 71} (2005) 049901]
  [arXiv:hep-th/0402183].
%
\end{thebibliography}
\end{document}